\documentclass[aps, prl, reprint,groupedaddress,footinbib, showpacs]{revtex4-1}

\usepackage{amsmath}
\usepackage{amssymb}
\usepackage{mathtools}
\usepackage{xcolor}
\usepackage{graphicx}
\usepackage{xfrac}
\usepackage{xcolor}
\usepackage{subfigure}
\usepackage{bm}
\usepackage{hyperref,hypcap}
\usepackage{ulem}

\begin{document}

\title{Electrical Reservoirs for Bilayer Excitons}
\author{Ming Xie}
\affiliation{Department of Physics, The University of Texas at Austin, Austin, TX 78712, USA}
\author{A. H. MacDonald}
\affiliation{Department of Physics, The University of Texas at Austin, Austin, TX 78712, USA}

\date{\today}

\begin{abstract}
	The ground state of two-dimensional (2D) electron systems with equal low 
	 densities of electrons and holes in nearby layers is an exciton fluid.  
	 We show that a reservoir for excitons can be established by 
	 contacting the two layers separately and maintaining the chemical
	 potential difference at a value less than the spatially indirect band gap, 
	 thereby avoiding the presence of free carriers in either layer.
	 Equilibration between the exciton fluid and the contacts proceeds via a process
	 involving virtual intermediate states in which an unpaired electron or hole virtually occupies a 
	 free carrier state in one of the 2D layers.  We derive an approximate 
	 relationship between the exciton-contact equilibration rate and 
	 the electrical conductances between the contacts and individual 2D layers when the contact chemical 
	 potentials align with the free-carrier bands, and explain how electrical measurements can be used to 
         measure thermodynamic properties of the exciton fluids.
\end{abstract}


\maketitle

\textit{Introduction.}---Excitons are composite bosonic particles in which conduction band electrons bind with 
valence band holes.  Excitons normally exist as
excited states of semiconductors and insulators, and
can have extremely long lifetimes when the electron and hole are 
separated in momentum-space, or in real-space \cite{Lozovik1976}, or both.  
Bose-Einstein condensation of long-lived excitons was predicted several 
decades ago \cite{Keldysh1968}, and is thought to have been realized relatively 
recently in semiconductor 
quantum well \cite{Butov2001, Butov2002, Gorbunov2006, High2012a, High2012b} double-layers.
Closely related polariton condensate states, in which longer range coherence is assisted
by the small masses of 2D vertical cavity photons, are regularly 
realized and have been studied extensively over the past decade  \cite{Deng2002,Kasprzak2006,Balili2007,Baumberg2008,Wertz2010,Snoke2002,Deng2010,Carusotto2013,Byrnes2014,Sanvitto2016,Sun2017}.
In typical exciton-condensation experiments a population of electrons and holes is generated in nearby 2D layers  
by optical excitation.  Free electrons and holes can also be injected electrically if contacts can be established to 
conduction and valence bands \cite{Schneider2013,Bhattacharya2013,Bhattacharya2014, Yao2012}.
The electrons and holes then combine to form excitons and the 
excitonic state is revealed by photons emitted during the exciton radiative decay process
\cite{Note1, Cohen2016, Shilo2013}.
\footnotetext[1]{For systems in which dark excitons (excitons with momentum or spin quantum numbers that
don't match those of the photons) are favored however, direct optical access is blocked.}
In this paper we propose and theoretically analyze a mechanism 
that allows direct electrical control of the chemical potential of spatially indirect exciton fluids
without populating free electron and hole states.
The mechanism requires substantial exciton binding energies in systems with 
long electron-hole recombination times.  Our proposal is motivated 
by the properties of van der Waals (vdW) heterojunction systems in which single-layer 
semiconductors are separated by hexagonal boron nitride (hBN) barrier layers.

\begin{figure}[h!]
	\includegraphics[width=0.45\textwidth]{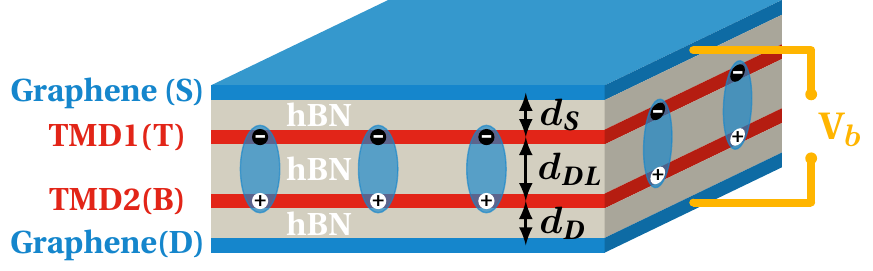}%
	\caption{\label{Fig:structure} (Color online) Schematic illustration of a 2D material heterojunction
	capable of supporting a spatially indirect exciton condensate, 
	and of an electrode pair that can act as a reservoir for spatially indirect excitons.}
\end{figure}

In recent years, 2D transition metal dichalcogenide (TMD) semiconductors have been established 
as an exciting exciton physics platform \cite{Qiu2013, Mak2013, Fogler2014} in which energy scales are enhanced by strong 
Coulomb interactions.  Surprising flexibility in the design of optical and electronic properties can be 
achieved \cite{Geim2013, Novoselov2016, Mak2016} by stacking vdW coupled 
2D materials in a variety of different arrangements.   
vdW heterojunctions involving 2D semiconductors can host spatially indirect excitons formed from 
electrons and holes in two different layers with binding energies
that remain large, even when the electron and hole layers are separated by 
hBN layers that increase 
exciton lifetimes by orders of magnitude from the nanosecond range \cite{Rivera2015, Palummo2015, Calman2017}
that applies in the absence of spacer layers. 

When the chemical potentials of contact electrodes
are inside the energy gaps of the 2D semiconductors,
electrons cannot tunnel into double-layer band states.
However, because of the Coulomb interaction and the exciton binding 
energy that it produces, correlated pair tunneling from electrodes 
connected to the two different layers is possible.  
In this Letter, we develop a microscopic model of this two-particle tunneling process and
argue that it can allow electrode pairs to act directly as exciton reservoirs
with a well-defined chemical potential set by the source-to-drain bias.
Direct exciton reservoirs have advantages for exciton generation and control over
the commonly employed indirect optical and electrical generation processes that start by generating free electrons and holes,
and we expect in particular that they will 
enable electrical measurements of the transport properties of exciton fluids.  

\begin{figure}
	\centering
	\includegraphics[width=0.49\textwidth]{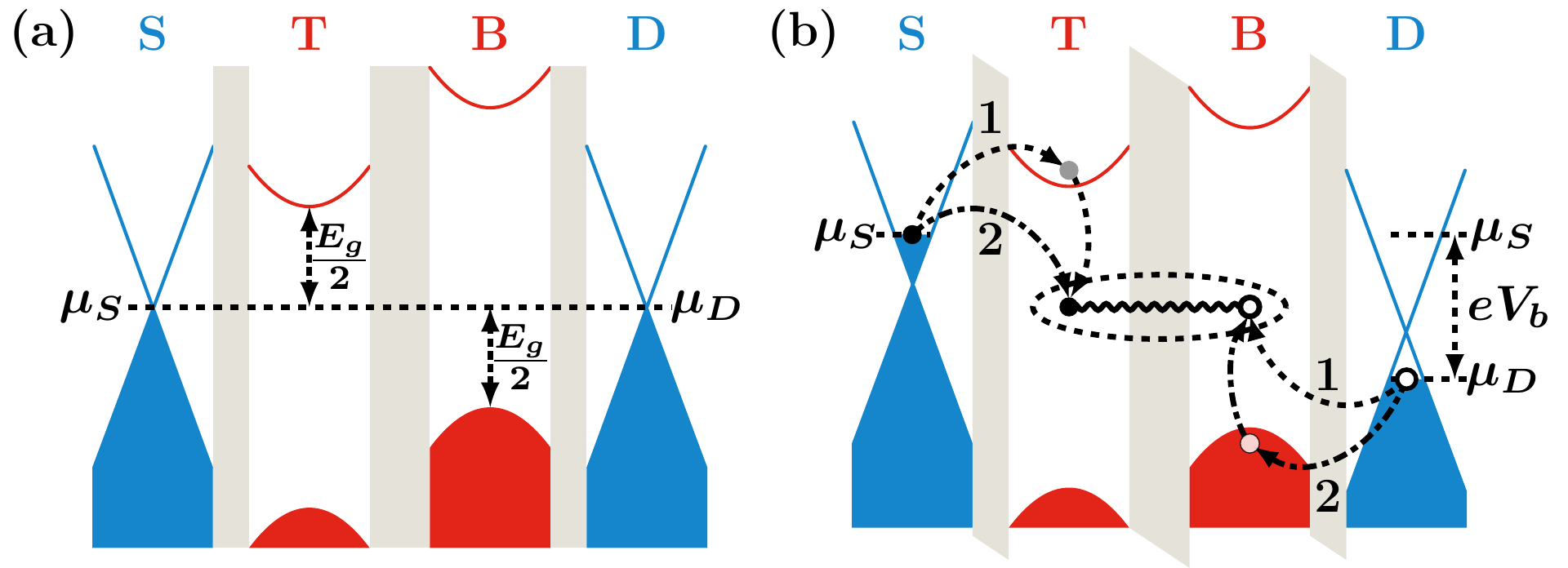}%
	\caption{ (Color online)\label{Fig:bands} 
		Schematic band diagrams for the vertical vdW heterostructure systems of interest
		for the case of (a) zero applied bias and (b) a finite bias $V_b$ satisfying 
		indirect gap $E_g > eV_b> \mu^0_{ex}$ where $\mu^0_{ex}$ is the 
		energy of an isolated spatially indirect exciton.  
		In case (b), the filled (empty) black circles represents electrons (holes) 
		and the gray (pink) circle represents a virtual state in the T(B) layer. 
		1 and 2 label two possible tunneling paths as discussed in detail in the text.}
\end{figure}

\textit{Correlated pair tunneling.}---We consider the vertically stacked multilayer heterostructure system illustrated 
in  Fig.~\ref{Fig:structure}, which contains a TMD semiconductor double-layer (DL) 
with a hBN barrier layer sufficiently thick to suppress tunneling, and 
source (S) and drain (D) electrodes that contact the two layers separately.
For the sake of definiteness we have assumed that the electrodes are formed from 
graphene sheets instead of metals since these have less influence on 
exciton binding energies \cite{raja2017}, but this detail is inessential.  
Dynamic screening due to source-drain electron-hole pair excitations is negligible
because of the suppression of source-to-drain tunneling by the tunnel barrier.
We also assume that the top (T) layer and bottom (B) layer materials are chosen so that the
conduction band minimum and valence band maximum are respectively above and below but 
close to the graphene sheet Dirac point (for example, T=MoS$_2$ and B=WSe$_2$ \cite{Zhang2017})
as illustrated in Fig.~\ref{Fig:bands}(a). 
Once the bias voltage between S and D, $\mu_S - \mu_D = eV_b$,
exceeds the energy needed to create an isolated indirect exciton, 
an exciton fluid will form whose equilibrium chemical potential equals $eV_b $, where $e>0$.
We discuss the equilibration process below. 

The total Hamiltonian of the four-layer system is
\begin{align}
\hat{\mathcal{H}} =
\hat{\mathcal{H}}_S +\hat{ \mathcal{H}}_D + 
\hat{\mathcal{H}}_{DL}+ \hat{\mathcal{H}}_t.
\end{align}
where $\hat{\mathcal{H}}_S$ and $\hat{\mathcal{H}}_D$ are the 
linear band Hamiltonians of the graphene electrodes, 
and $\hat{\mathcal{H}}_{DL}$ is the DL Hamiltonian including Coulomb interactions.
In this paper we assume that the TMD DL is in its exciton 
condensate ground state and ignore its spin degree of freedom. 
Tunneling between the electrodes and the double-layer system is accounted for by 
\begin{align}
\hat{\mathcal{H}}_t&=
\sum_{\bm{k},\bar{\bm{p}}}
t^S_{\bm{k} \bar{\bm{p}}} \,
\hat{c}^{\dagger}_{\bm{k},T}
\hat{a}^{\vphantom{\dagger}}_{\bar{\bm{p}},S}
+ t^{D}_{\bm{k} \bar{\bm{p}}} \, 
\hat{c}^{\dagger}_{\bm{k},B}
\hat{a}^{\vphantom{\dagger}}_{\bar{\bm{p}},D}
+h.c. \label{ht}
\end{align}
where $t^{S(D)}_{\bm{k} \bar{\bm{p}}}$ is a tunneling matrix element.
$\bar{\bm{p}} \equiv (\bm{p},\lambda, \tau)$ is a compound index 
that combines the 2D momentum $\bm{p}$, 
the band index $\lambda=c,v$ and the valley index $\tau$ of the graphene electrode states.
In Eq.~(\ref{ht}), $\hat{a}$ is the creation operator in the electrodes and $\hat{c}^\dagger_{T(B)}$ is the creation 
operator for conduction band electrons in T and valence band electrons in B.  
For single-grain hBN tunnel barriers, the tunneling properties can 
have very specific momentum dependence which does not play 
an essential role and is not accounted for below, but is sensitive to the relative orientation of the 
various 2D material layers \cite{Bistritzer2010, Zhou2017}.
We neglect interlayer tunneling between the T and B primarily because 
we are interested in a bias voltage regime in which free carriers are not present to tunnel.  
We also set the interlayer radiative recombination rate to zero in order to 
focus on double-electrode reservoir properties.  
In practice we anticipate that the interplay between our 
exciton reservoirs and interlayer radiative recombination, whose strength can be adjusted over orders of magnitude by varying the thickness and orientation of the hBN barrier layer, opens up a rich range of opto-electronic 
phenomena for study that are a primary motivation for this work.  

The band diagram of the vertical vdW heterostructure system is shown schematically in Fig.~\ref{Fig:bands}. 
At zero bias (Fig.~\ref{Fig:bands}(a)), we assume that both graphene 
electrodes are neutral and that the aligned Dirac points are
in the middle of the spatially indirect band gap $E_{g}$.
When a bias voltage in the subgap regime ($E_g> eV_b >0$) is applied,
tunneling between the electrodes and free-carrier states in the TMD layers is prohibited by energy conservation.
(Note that $E_g$ increases with $V_b$, and that direct tunneling of electrons 
from source to drain is extremely strongly suppressed because it must navigate three tunneling barriers.)  
Our interest here is in the bias regime $E_g>V_b>\mu^0_{ex}$,
where $\mu^0_{ex}$ is the energy of an isolated spatially indirect exciton.  
In this bias voltage regime energy conservation can be achieved by 
the two-particle tunneling process
illustrated in Fig.~ \ref{Fig:bands}(b). 
The state created when an electron from S tunnels to a virtual state in T and a hole from D subsequently tunnels to B has a finite overlap with an exciton fluid state (path 1). 
An alternative and equally possible path is for a hole from D to 
tunnel to a virtual state in B first (path 2).
Each tunneling process effectively transfers one electron from S to D and creates an exciton in the DL. 
We concentrate here on the case of $T<T_c$ 
for which the excitons form a condensate, although the main idea of 
using an electrode pair as a reservoir for excitons applies equally well when the 
excitons form a non-condensed gas.  
For the low temperature case we find that because of the stimulated scattering characteristic of bosonic 
statistics, a major fraction of the excitons added or removed from the 
system are simply added or removed from the condensate.

The condensate state of spatially indirect excitons has been 
extensively studied in previous work \cite{Comte1982, Zhu1995, Wu2015} 
using a BCS-like mean field theory approach
in which the ground state is found by minimizing 
$\langle \hat{\mathcal{H}}_{DL}-\mu_{ex}\hat{\mathcal{N}}\rangle$ in the space of 
Slater determinant states with coherence between conduction and valence bands.   
Here $\hat{\mathcal{N}} = \sum_{\bm{k}}( \hat{c}^{\dagger}_{\bm{k},T}\hat{c}^{\vphantom{\dagger}}_{\bm{k},T} +
\hat{c}^{\vphantom{\dagger}}_{\bm{k},B} \hat{c}^{\dagger}_{\bm{k},B}  )/2$ 
is the total number of electron-hole pairs.
The mean-field Hamiltonian of the exciton condensate system is 
\begin{align}
\hat{\mathcal{H}}^{MF}_{DL} = E_G+ \sum_{\mathclap{\bm{k}}} 
E_{\bm{k}}
(
\hat{\gamma}^{\dagger}_{\bm{k}, 0} \hat{\gamma}^{\vphantom{\dagger}}_{\bm{k}, 0} -
\hat{\gamma}^{\dagger}_{\bm{k}, 1}  \hat{\gamma}^{\vphantom{\dagger}}_{\bm{k}, 1}
)
+
\mu_{ex}\hat{\mathcal{N}}
\end{align}
where $E_G$ is the condensate ground state energy and $\mu_{ex}$ the exciton chemical potential. The quasiparticle energy dispersion is $E_{\bm{k}}=\sqrt{ (\epsilon^T_{\bm{k}}-\mu_{ex}/2)^2 + \Delta_{\bm{k}}^2 },$
where we have assumed that the DL energy dispersion $\epsilon^T_{\bm{k}}=-\epsilon^B_{\bm{k}}=\hbar^2k^2/(2m)+E_g/2$ and
that the order parameter $\Delta_{\bm{k}}$ is real. 
 $\hat{\gamma}^{\dagger}_{\bm{k}, 0} = u_{\bm{k}} \hat{c}^{\dagger}_{\bm{k}, T} + v_{\bm{k}} \hat{c}^{\dagger}_{\bm{k},B}$ 
 and 
 $\gamma^{\dagger}_{\bm{k}, 1} = v_{\bm{k}} \hat{c}^{\dagger}_{\bm{k}, T} - u_{\bm{k}} \hat{c}^{\dagger}_{\bm{k},B}$ 
 are creation operators for states in the empty and occupied dressed quasiparticle bands, respectively,
and $u_{\bm{k}}$ and $v_{\bm{k}}$ are coherence factors that depend on the pair potential $\Delta_{\bm{k}}$ 
which is determined in turn by solving a self-consistent equation that has solutions 
only if $\mu_{ex}$ exceeds $\mu^0_{ex}$ \cite{Comte1982}.  

The two-particle tunneling rate can be obtained
by applying Fermi's golden rule to the second order
tunneling process.  We find that 
the net rate at which excitons are added to the condensate is   
\begin{align}
\label{eq:rate}
\frac{dn_{ex}}{dt}
=
\frac{2\pi}{\hbar A} \sum_{\bar{\bm{p}},\bar{\bm{p}}'}
\left| M_{\bar{\bm{p}}\bar{\bm{p}}'}\right|^2
(f_{\bar{\bm{p}}}^S - f_{\bar{\bm{p}}'}^D)
\delta(\epsilon^S_{\bar{\bm{p}}} - \epsilon^D_{\bar{\bm{p}}'} - \mu_{ex}  )
\end{align}
where $f^{\alpha}_{\bar{\bm{p}}}$, with $\alpha=S, D$, is the Fermi distribution function.
The matrix element in Eq.~(\ref{eq:rate}) 
\begin{align}
	M_{ \bar{\bm{p}} \bar{\bm{p}}'} = \sum_{\bm{k}} u_{\bm{k}} v_{\bm{k}} t^S_{\bm{k} \bar{\bm{p}}  } t^{D*}_{ \bm{k} \bar{\bm{p}}' }
	\left\{
	  \frac{1}{ E^0_{\bm{k}} - \epsilon_{\bar{\bm{p}}}^S } + 
	  \frac{1}{ \epsilon_{\bar{\bm{p}}'}^D - E^1_{\bm{k}} }
	\right\} \label{melement}
\end{align}
where $E^0_{\bm{k}}=E_{\bm{k}} + \mu_{ex}/2$ and $E^1_{\bm{k}}=-E^0_{\bm{k}}$ are the energies required
to add quasiparticles of momentum $\bm{k}$ to bands $0$ and $1$, respectively.
The energy denominators account for the finite energy cost of hopping to the intermediate virtual states, 
and never vanish in the bias voltage range of interest.  
The two terms in the matrix element account for the two tunneling paths
depicted in Fig.~\ref{Fig:bands}(b).

\begin{figure}
	\centering
	\includegraphics[width=0.475\textwidth]{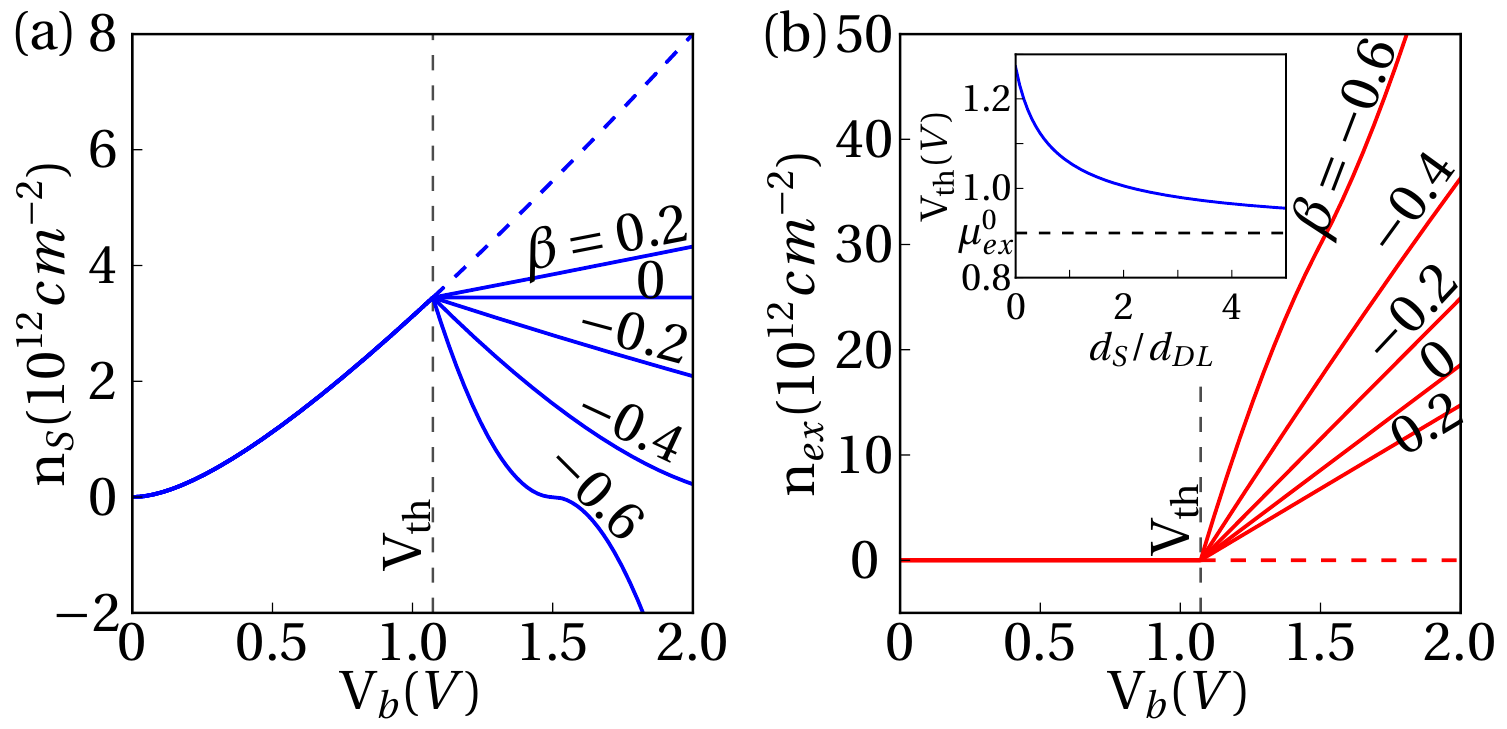}%
	\caption{\label{Fig:dcplot}Equilibrium electrode densities (a) and exciton densities (b) 
		at different values of $\beta = g_{XC}/g_H$, where $g_H$ is fixed with its value set by
		choosing $d_{DL}=1nm$.   These results were
		calculated for $d_S=d_D=d_{DL}$ spatially indirect gap $E_g(0)=1.1eV$,
		and exciton binding energy $E_b=0.2eV$.
		The extended (colored) dashed line represents electrode densities in the case in 
		which the TMD double-layer that hosts excitons is absent.
		The vertical gray dashed lines indicates the threshold voltage at which 
		the dual electrodes establish a reservoir for excitons.
		Inset: Threshold voltage as a function of $d_S/d_{DL}$.  
		The dashed line indicates the zero electrode density limit of the isolated  
		exciton energy $\mu^0_{ex}$.}
\end{figure}

The evaluation of $ \left| M_{ \bar{\bm{p}} \bar{\bm{p}}'} \right|^2 $ in Eq.~(\ref{melement}) requires
 knowledge of the momentum dependent tunneling amplitudes $t^{\alpha}_{\bm{k}\bar{\bm{p}}}$.
We simplify our calculation by assuming that 
interfacial disorder plays an important role in determining the tunneling amplitude.
We employ a Gaussian random tunneling model for which 
$\langle t^{\alpha}(\bm{r}) \rangle_{\text{dis}} = 0$ and 
the second order correlation functions satisfy
\begin{align}
	&\langle t^{\alpha}(\bm{r}) t^{\alpha'*}(\bm{r}') \rangle_{\text{dis}} = |\Delta t|^2 \mathcal{F}( \bm{r} - \bm{r}' )
	\delta_{\alpha\alpha'},
\end{align}
where $\langle \cdots \rangle_{\text{dis}}$ is the disorder average and $\mathcal{F}( \bm{r} - \bm{r}' )$ is a smoothly decaying function of the distance $|\bm{r} - \bm{r}'|$.
For low exciton densities and $V_{b}>\mu^0_{ex}$ limit, we obtain the tunneling current-voltage equation
\begin{align}
	I_{ex} \approx G_{ex}(V_b-\mu_{ex}/e).
	\label{ivrelation}
\end{align}
($\mu_{ex} > \mu^0_{ex}$ at finite exciton density.)
The effective exciton tunneling conductance $G_{ex}$ 
is given approximately by 
\begin{align}
	G_{ex}=\frac{A g^{S}_N g^{D}_Nn_{ex}a_B^2}{e^2/\hbar}\frac{8}{\rho_0 E_b},
	\label{tunnelcond}
\end{align}
where $g^{S}_N$ and $g^{D}_N$ are the normal tunneling conductances per unit area 
between S and T and between D and B, respectively,
$\rho_0$ is the density of states of quasiparticle band $0$,
$a_B$ is the Bohr radius, and $E_b$ the exciton binding energy.
The tunneling conductance in 
Eq.~(\ref{tunnelcond}) is proportional to the exciton condensate 
density $n_{ex}$ because of the bosonic stimulated scattering effect.  
Since the fraction of uncondensed excitons is small in the low density BEC limit, 
we have assumed in deriving this simple result that the contributions from processes with a final state 
exciton outside the condensate are negligible.
Using the typical values 
$g_N^S=g_N^D \sim 10^{-2} e^2/h \cdot \mu m^{-2}$, $n_{ex}a_B^2 \sim 0.01$,
and taking the quasiparticle band masses
in the TMD layers close to the bare electron mass, we estimate that 
$G_{ex}$ is in the order $10^{-11} e^2/h \cdot \mu m^{-2}$.

\begin{figure}[h]
	\centering
	\includegraphics[width=0.5\textwidth]{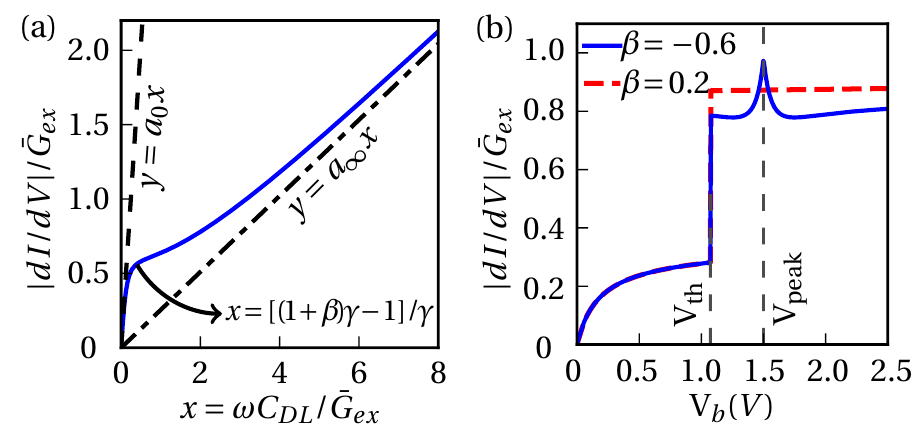}
	\caption{\label{Fig:acplot} (Color online) 
		Differential conductance as a function of (a) $x=\omega C_{DL}/\bar{G}_{ex}$ and (b) bias voltage $V_b$. 
		The dashed line and the dash-dotted line in (a) shows the linear relation at low and high frequency limits, respectively.}
\end{figure}

Eq.~(\ref{ivrelation}) states that a time-independent quasi-equilibrium is reached when 
$eV_b=\mu_{ex}$.  We say quasi-equilibrium here,
rather than simply equilibrium, to emphasize that we are assuming that excitons 
cannot annihilate by radiative recombination.
As long as all processes in which electrons move between the two
TMD layers are absent, we effectively have an equilibrium problem in 
which the spatially indirect band gap is tuned electrically by varying $V_{b}$.  
We do not emphasize this distinction between equilibrium and 
quasi-equilibrium below. 

\textit{Electrical characteristics of exciton reservoirs.}---Because of repulsive interactions between excitons,
their chemical potential increases with exciton density \cite{Comte1982, Zhu1995, Wu2015}.
For spatially indirect excitons $\mu_{ex} = \mu^{0}_{ex} +  (g_H + g_{XC}) n_{ex}$,
where $g_H =  e^2/C_{DL} = \epsilon/(4\pi d_{DL})$ and $g_{XC}$ are the Hartree and 
exchange-correlation contributions to the effective exciton-exciton interaction.
The Hartree term accounts for the capacitive coupling between T and B layers,
whereas $g_{XC}$, which is density-dependent and cannot 
be evaluated exactly, accounts for exchange and correlation effects 
contributions due to both intralayer and interlayer Coulombic interactions and  
therefore also depend on the interlayer spacing $d_{DL}$.
When we add the potential energy associated
with charged electrodes, the exciton
chemical potential in our geometry has an additional electrostatic contribution: 
\begin{align}
	\mu_{ex} = \mu^0_{ex} +  (g_H + g_{XC}) n_{ex} + g_H n_S  \label{muex}
\end{align}
where $n_S = n_D $ are the equal densities of electrons and 
holes in the two electrodes. 
Eq.~(\ref{muex}) can be obtained by minimizing the 
total energy, $\mathcal{E}[n_S,n_{ex}]$, 
with respect to $n_{ex}$ 
(see Supplemental Material
\setcounter{footnote}{16}\footnote{See Supplemental Material for the discussion of the total energy functional $\mathcal{E}[n_S,n_{ex}]$,  the existence of a global minimum and a detailed derivation of the {\it ac} response equations.}). 
By minimizing $\mathcal{E}[n_S,n_{ex}]$ with respect to 
$n_{S}$ we obtain the following expression for the bias potential, 
\begin{align}
&\frac{n_{ex}}{C_{DL}}
+
n_S \left( \frac{2 \hbar v \sqrt{2\pi}}{e^2 \sqrt{|n_S|}} + \frac{1}{C_{geo}}\right) = \frac{V_b}{e}.
\label{staticeq}
\end{align}
In Eq.~(\ref{staticeq}), $C_{geo} = \epsilon/ (4 \pi e^2 d_{tot}) $ is the geometric capacitance and 
$d_{tot} = d_{S} + d_{DL} +d_{D}$; the energy function used to derive Eq.~(\ref{staticeq})
does not account for exchange and correlation in the electrodes, {\it i.e.} for interaction corrections to the 
electrode quantum capacitance, but these can easily be added when relevant.
A time-independent equilibrium between the electrodes and the exciton fluid
is established when $n_{ex} \ne 0$ and electron-hole pairs 
have the same chemical potential in either environment 
{\it i.e.} when  and $eV_b=\mu_{ex}$.


Fig.~\ref{Fig:dcplot} shows equilibrium densities calculated
for several typical values of the dimensionless exchange-correlation
coupling strength $\beta \equiv g_{XC}/g_H$.
Below we take $\beta$ as an unknown parameter and show that its value
can be measured electrically.  When estimated using self-consistent mean-field theory 
$\beta$ changes sign from positive to negative when $d_{DL}$ exceeds around a quarter of 
an excitonic Bohr radius, and using TMD semiconductor parameters 
has the value $\beta=-0.6$ for 
$d_{DL}=1$nm in \cite{Wu2015}.
Below the threshold voltage $V_{\rm th}$,
which satisfies $eV_{\rm th}=\mu^0_{ex}+g_H n_S(V_{\rm th})$ and depends on $d_S/d_{DL}$,
no excitons are injected and $n_S(V_{b})$ is independent of $\beta$, 
as shown in Fig.~{\ref{Fig:dcplot}}(a).
When $V_b>V_{\rm th}$, electrons and holes can enter the TMD layers by 
forming excitons via the two-particle tunneling process.
The slope of the $n_S(V_{b})$ curve is reduced and becomes 
negative when $\beta$ changes sign from positive to negative.
For $\beta<0$, we find that $n_S$ becomes negative
when $V_b=-\mu^0_{ex}/\beta$; we show later that 
the dynamic response is anomalous at this point.

The two-particle tunneling rate which we have estimated theoretically
can be measured by performing {\it ac} electrical measurements, letting  
$V_b(t)=V_{dc}+V_{ac}\cos{\omega t}$, where $V_{ac}$ is small.
The linear response of the system to $V_{ac}$ 
can be extracted by 
linearizing Eq.~(\ref{ivrelation}) and (\ref{staticeq}) 
(see details in Supplemental Material \cite{Note17}).
In Fig.~{\ref{Fig:acplot}} we plot
normalized amplitudes of the differential conductance $|dI/dV|/\bar{G}_{ex}$, where 
$\bar{G}_{ex}(V_{dc})$ is the $\it{dc}$ exciton conductance $G_{ex}$.
In the low and high frequency limits, the system behaves effectively as a capacitor with
$C^{eff}_{low}=a_0C_{DL}$, $C^{eff}_{high}=a_{\infty}C_{DL}$, 
where $\gamma=(1+2C_{geo}/C_Q)d_{tot}/d_{DL}$ and 
$a_0=\{(\gamma-1)^2/[(1+\beta)\gamma-1]+1\}/\gamma$ and $a_{\infty}=1/\gamma$.
$C_Q=\sqrt{2n_S}/(\sqrt{\pi}\hbar v)$ is the quantum capacitance of graphene \cite{Xia2009}.
$dI/dV$ deviates from linear frequency dependence when the 
scaled frequency $x=\omega C_{DL}/\bar{G}_{ex} \sim x_0=[(1+\beta)\gamma-1]/\gamma$.
Measuring the crossover frequency $\omega_0$, gives the tunneling conductance 
$\bar{G}_{ex}=\omega_0C_{DL}/x_0$.
The {\it dc} bias voltage dependence of the differential conductance is shown in Fig.~{\ref{Fig:acplot}}(b).
For $\beta<0$ a peak appears at $V_{\rm peak}=-E^0_{ex}/\beta$, the point at which $n_S$ reaches $0$, 
as mentioned above.  This suggests a way to measure the exciton exchange-correlation
energy parameter $g_{XC}$, provided that $E_g$ and $E_b$ are known.
For $\beta>0$, the differential conductance increases slowly with increasing 
$V_b$ in the regime $V_b>V_{\rm th}$. 

Optical recombination, which provides a mechanism 
for excitons to leak out of the system of interest, can easily be added to the 
theory explained here, and in the {\em dc} bias voltage case 
converts the equilibrium exciton fluid into a steady state.  When the steady
state exciton fluid condenses, it will emit coherent light forming a state similar to
a polariton laser but subject to precise electrical control.


This work was supported by the Army Research Office 
under Award W911NF-17-1-0312 
and by the Welch Foundation under grant TBF1473.
Ming Xie was supported by The Center for Dynamics and Control of Materials (CDCM) 
under NSF Award DMR-1720595.
The authors acknowledge helpful discussions with Hui Deng,
Jim Eisenstein, Emanuel Tutuc, and David Snoke.  

\bibliography{electrical_exciton}{}

\begin{thebibliography}{43}%
\makeatletter
\providecommand \@ifxundefined [1]{%
 \@ifx{#1\undefined}
}%
\providecommand \@ifnum [1]{%
 \ifnum #1\expandafter \@firstoftwo
 \else \expandafter \@secondoftwo
 \fi
}%
\providecommand \@ifx [1]{%
 \ifx #1\expandafter \@firstoftwo
 \else \expandafter \@secondoftwo
 \fi
}%
\providecommand \natexlab [1]{#1}%
\providecommand \enquote  [1]{``#1''}%
\providecommand \bibnamefont  [1]{#1}%
\providecommand \bibfnamefont [1]{#1}%
\providecommand \citenamefont [1]{#1}%
\providecommand \href@noop [0]{\@secondoftwo}%
\providecommand \href [0]{\begingroup \@sanitize@url \@href}%
\providecommand \@href[1]{\@@startlink{#1}\@@href}%
\providecommand \@@href[1]{\endgroup#1\@@endlink}%
\providecommand \@sanitize@url [0]{\catcode `\\12\catcode `\$12\catcode
  `\&12\catcode `\#12\catcode `\^12\catcode `\_12\catcode `\%12\relax}%
\providecommand \@@startlink[1]{}%
\providecommand \@@endlink[0]{}%
\providecommand \url  [0]{\begingroup\@sanitize@url \@url }%
\providecommand \@url [1]{\endgroup\@href {#1}{\urlprefix }}%
\providecommand \urlprefix  [0]{URL }%
\providecommand \Eprint [0]{\href }%
\providecommand \doibase [0]{http://dx.doi.org/}%
\providecommand \selectlanguage [0]{\@gobble}%
\providecommand \bibinfo  [0]{\@secondoftwo}%
\providecommand \bibfield  [0]{\@secondoftwo}%
\providecommand \translation [1]{[#1]}%
\providecommand \BibitemOpen [0]{}%
\providecommand \bibitemStop [0]{}%
\providecommand \bibitemNoStop [0]{.\EOS\space}%
\providecommand \EOS [0]{\spacefactor3000\relax}%
\providecommand \BibitemShut  [1]{\csname bibitem#1\endcsname}%
\let\auto@bib@innerbib\@empty
\bibitem [{\citenamefont {Lozovik}\ and\ \citenamefont
  {Yudson}(1976)}]{Lozovik1976}%
  \BibitemOpen
  \bibfield  {author} {\bibinfo {author} {\bibfnamefont {Y.}~\bibnamefont
  {Lozovik}}\ and\ \bibinfo {author} {\bibfnamefont {V.}~\bibnamefont
  {Yudson}},\ }\href {http://www.jetpletters.ac.ru/ps/1530/article_23399.shtml}
  {\bibfield  {journal} {\bibinfo  {journal} {JETP Lett.}\ }\textbf {\bibinfo
  {volume} {22}},\ \bibinfo {pages} {274} (\bibinfo {year} {1976})}\BibitemShut
  {NoStop}%
\bibitem [{\citenamefont {{Keldysh}}\ and\ \citenamefont
  {{Kozlov}}(1968)}]{Keldysh1968}%
  \BibitemOpen
  \bibfield  {author} {\bibinfo {author} {\bibfnamefont {L.~V.}\ \bibnamefont
  {{Keldysh}}}\ and\ \bibinfo {author} {\bibfnamefont {A.~N.}\ \bibnamefont
  {{Kozlov}}},\ }\href
  {http://www.jetp.ac.ru/cgi-bin/e/index/e/27/3/p521?a=list} {\bibfield
  {journal} {\bibinfo  {journal} {Sov. Phys. JETP}\ }\textbf {\bibinfo {volume}
  {27}},\ \bibinfo {pages} {521} (\bibinfo {year} {1968})}\BibitemShut
  {NoStop}%
\bibitem [{\citenamefont {Butov}\ \emph {et~al.}(2001)\citenamefont {Butov},
  \citenamefont {Ivanov}, \citenamefont {Imamoglu}, \citenamefont {Littlewood},
  \citenamefont {Shashkin}, \citenamefont {Dolgopolov}, \citenamefont
  {Campman},\ and\ \citenamefont {Gossard}}]{Butov2001}%
  \BibitemOpen
  \bibfield  {author} {\bibinfo {author} {\bibfnamefont {L.~V.}\ \bibnamefont
  {Butov}}, \bibinfo {author} {\bibfnamefont {A.~L.}\ \bibnamefont {Ivanov}},
  \bibinfo {author} {\bibfnamefont {A.}~\bibnamefont {Imamoglu}}, \bibinfo
  {author} {\bibfnamefont {P.~B.}\ \bibnamefont {Littlewood}}, \bibinfo
  {author} {\bibfnamefont {A.~A.}\ \bibnamefont {Shashkin}}, \bibinfo {author}
  {\bibfnamefont {V.~T.}\ \bibnamefont {Dolgopolov}}, \bibinfo {author}
  {\bibfnamefont {K.~L.}\ \bibnamefont {Campman}}, \ and\ \bibinfo {author}
  {\bibfnamefont {A.~C.}\ \bibnamefont {Gossard}},\ }\href {\doibase
  10.1103/PhysRevLett.86.5608} {\bibfield  {journal} {\bibinfo  {journal}
  {Phys. Rev. Lett.}\ }\textbf {\bibinfo {volume} {86}},\ \bibinfo {pages}
  {5608} (\bibinfo {year} {2001})}\BibitemShut {NoStop}%
\bibitem [{\citenamefont {Butov}\ \emph {et~al.}(2002)\citenamefont {Butov},
  \citenamefont {Lai}, \citenamefont {Ivanov}, \citenamefont {Gossard},\ and\
  \citenamefont {Chemla}}]{Butov2002}%
  \BibitemOpen
  \bibfield  {author} {\bibinfo {author} {\bibfnamefont {L.~V.}\ \bibnamefont
  {Butov}}, \bibinfo {author} {\bibfnamefont {C.~W.}\ \bibnamefont {Lai}},
  \bibinfo {author} {\bibfnamefont {A.~L.}\ \bibnamefont {Ivanov}}, \bibinfo
  {author} {\bibfnamefont {A.~C.}\ \bibnamefont {Gossard}}, \ and\ \bibinfo
  {author} {\bibfnamefont {D.~S.}\ \bibnamefont {Chemla}},\ }\href
  {http://dx.doi.org/10.1038/417047a} {\bibfield  {journal} {\bibinfo
  {journal} {Nature}\ }\textbf {\bibinfo {volume} {417}},\ \bibinfo {pages}
  {47} (\bibinfo {year} {2002})}\BibitemShut {NoStop}%
\bibitem [{\citenamefont {Gorbunov}\ and\ \citenamefont
  {Timofeev}(2006)}]{Gorbunov2006}%
  \BibitemOpen
  \bibfield  {author} {\bibinfo {author} {\bibfnamefont {A.~V.}\ \bibnamefont
  {Gorbunov}}\ and\ \bibinfo {author} {\bibfnamefont {V.~B.}\ \bibnamefont
  {Timofeev}},\ }\href {\doibase 10.1134/S0021364006040047} {\bibfield
  {journal} {\bibinfo  {journal} {JETP Lett.}\ }\textbf {\bibinfo {volume}
  {83}},\ \bibinfo {pages} {146} (\bibinfo {year} {2006})}\BibitemShut
  {NoStop}%
\bibitem [{\citenamefont {High}\ \emph
  {et~al.}(2012{\natexlab{a}})\citenamefont {High}, \citenamefont {Leonard},
  \citenamefont {Hammack}, \citenamefont {Fogler}, \citenamefont {Butov},
  \citenamefont {Kavokin}, \citenamefont {Campman},\ and\ \citenamefont
  {Gossard}}]{High2012a}%
  \BibitemOpen
  \bibfield  {author} {\bibinfo {author} {\bibfnamefont {A.~A.}\ \bibnamefont
  {High}}, \bibinfo {author} {\bibfnamefont {J.~R.}\ \bibnamefont {Leonard}},
  \bibinfo {author} {\bibfnamefont {A.~T.}\ \bibnamefont {Hammack}}, \bibinfo
  {author} {\bibfnamefont {M.~M.}\ \bibnamefont {Fogler}}, \bibinfo {author}
  {\bibfnamefont {L.~V.}\ \bibnamefont {Butov}}, \bibinfo {author}
  {\bibfnamefont {A.~V.}\ \bibnamefont {Kavokin}}, \bibinfo {author}
  {\bibfnamefont {K.~L.}\ \bibnamefont {Campman}}, \ and\ \bibinfo {author}
  {\bibfnamefont {A.~C.}\ \bibnamefont {Gossard}},\ }\href
  {http://dx.doi.org/10.1038/nature10903} {\bibfield  {journal} {\bibinfo
  {journal} {Nature}\ }\textbf {\bibinfo {volume} {483}},\ \bibinfo {pages}
  {584} (\bibinfo {year} {2012}{\natexlab{a}})}\BibitemShut {NoStop}%
\bibitem [{\citenamefont {High}\ \emph
  {et~al.}(2012{\natexlab{b}})\citenamefont {High}, \citenamefont {Leonard},
  \citenamefont {Remeika}, \citenamefont {Butov}, \citenamefont {Hanson},\ and\
  \citenamefont {Gossard}}]{High2012b}%
  \BibitemOpen
  \bibfield  {author} {\bibinfo {author} {\bibfnamefont {A.~A.}\ \bibnamefont
  {High}}, \bibinfo {author} {\bibfnamefont {J.~R.}\ \bibnamefont {Leonard}},
  \bibinfo {author} {\bibfnamefont {M.}~\bibnamefont {Remeika}}, \bibinfo
  {author} {\bibfnamefont {L.~V.}\ \bibnamefont {Butov}}, \bibinfo {author}
  {\bibfnamefont {M.}~\bibnamefont {Hanson}}, \ and\ \bibinfo {author}
  {\bibfnamefont {A.~C.}\ \bibnamefont {Gossard}},\ }\href {\doibase
  10.1021/nl300983n} {\bibfield  {journal} {\bibinfo  {journal} {Nano Lett.}\
  }\textbf {\bibinfo {volume} {12}},\ \bibinfo {pages} {2605} (\bibinfo {year}
  {2012}{\natexlab{b}})}\BibitemShut {NoStop}%
\bibitem [{\citenamefont {Deng}\ \emph {et~al.}(2002)\citenamefont {Deng},
  \citenamefont {Weihs}, \citenamefont {Santori}, \citenamefont {Bloch},\ and\
  \citenamefont {Yamamoto}}]{Deng2002}%
  \BibitemOpen
  \bibfield  {author} {\bibinfo {author} {\bibfnamefont {H.}~\bibnamefont
  {Deng}}, \bibinfo {author} {\bibfnamefont {G.}~\bibnamefont {Weihs}},
  \bibinfo {author} {\bibfnamefont {C.}~\bibnamefont {Santori}}, \bibinfo
  {author} {\bibfnamefont {J.}~\bibnamefont {Bloch}}, \ and\ \bibinfo {author}
  {\bibfnamefont {Y.}~\bibnamefont {Yamamoto}},\ }\href {\doibase
  10.1126/science.1074464} {\bibfield  {journal} {\bibinfo  {journal}
  {Science}\ }\textbf {\bibinfo {volume} {298}},\ \bibinfo {pages} {199}
  (\bibinfo {year} {2002})}\BibitemShut {NoStop}%
\bibitem [{\citenamefont {Kasprzak}\ \emph {et~al.}(2006)\citenamefont
  {Kasprzak}, \citenamefont {Richard}, \citenamefont {Kundermann},
  \citenamefont {Baas}, \citenamefont {Jeambrun}, \citenamefont {Keeling},
  \citenamefont {Marchetti}, \citenamefont {Szymanska}, \citenamefont {Andre},
  \citenamefont {Staehli}, \citenamefont {Savona}, \citenamefont {Littlewood},
  \citenamefont {Deveaud},\ and\ \citenamefont {Dang}}]{Kasprzak2006}%
  \BibitemOpen
  \bibfield  {author} {\bibinfo {author} {\bibfnamefont {J.}~\bibnamefont
  {Kasprzak}}, \bibinfo {author} {\bibfnamefont {M.}~\bibnamefont {Richard}},
  \bibinfo {author} {\bibfnamefont {S.}~\bibnamefont {Kundermann}}, \bibinfo
  {author} {\bibfnamefont {A.}~\bibnamefont {Baas}}, \bibinfo {author}
  {\bibfnamefont {P.}~\bibnamefont {Jeambrun}}, \bibinfo {author}
  {\bibfnamefont {J.~M.~J.}\ \bibnamefont {Keeling}}, \bibinfo {author}
  {\bibfnamefont {F.~M.}\ \bibnamefont {Marchetti}}, \bibinfo {author}
  {\bibfnamefont {M.~H.}\ \bibnamefont {Szymanska}}, \bibinfo {author}
  {\bibfnamefont {R.}~\bibnamefont {Andre}}, \bibinfo {author} {\bibfnamefont
  {J.~L.}\ \bibnamefont {Staehli}}, \bibinfo {author} {\bibfnamefont
  {V.}~\bibnamefont {Savona}}, \bibinfo {author} {\bibfnamefont {P.~B.}\
  \bibnamefont {Littlewood}}, \bibinfo {author} {\bibfnamefont
  {B.}~\bibnamefont {Deveaud}}, \ and\ \bibinfo {author} {\bibfnamefont
  {L.~S.}\ \bibnamefont {Dang}},\ }\href
  {http://dx.doi.org/10.1038/nature05131} {\bibfield  {journal} {\bibinfo
  {journal} {Nature}\ }\textbf {\bibinfo {volume} {443}},\ \bibinfo {pages}
  {409} (\bibinfo {year} {2006})}\BibitemShut {NoStop}%
\bibitem [{\citenamefont {Balili}\ \emph {et~al.}(2007)\citenamefont {Balili},
  \citenamefont {Hartwell}, \citenamefont {Snoke}, \citenamefont {Pfeiffer},\
  and\ \citenamefont {West}}]{Balili2007}%
  \BibitemOpen
  \bibfield  {author} {\bibinfo {author} {\bibfnamefont {R.}~\bibnamefont
  {Balili}}, \bibinfo {author} {\bibfnamefont {V.}~\bibnamefont {Hartwell}},
  \bibinfo {author} {\bibfnamefont {D.}~\bibnamefont {Snoke}}, \bibinfo
  {author} {\bibfnamefont {L.}~\bibnamefont {Pfeiffer}}, \ and\ \bibinfo
  {author} {\bibfnamefont {K.}~\bibnamefont {West}},\ }\href {\doibase
  10.1126/science.1140990} {\bibfield  {journal} {\bibinfo  {journal}
  {Science}\ }\textbf {\bibinfo {volume} {316}},\ \bibinfo {pages} {1007}
  (\bibinfo {year} {2007})}\BibitemShut {NoStop}%
\bibitem [{\citenamefont {Baumberg}\ \emph {et~al.}(2008)\citenamefont
  {Baumberg}, \citenamefont {Kavokin}, \citenamefont {Christopoulos},
  \citenamefont {Grundy}, \citenamefont {Butt\'e}, \citenamefont {Christmann},
  \citenamefont {Solnyshkov}, \citenamefont {Malpuech}, \citenamefont
  {Baldassarri H\"oger~von H\"ogersthal}, \citenamefont {Feltin}, \citenamefont
  {Carlin},\ and\ \citenamefont {Grandjean}}]{Baumberg2008}%
  \BibitemOpen
  \bibfield  {author} {\bibinfo {author} {\bibfnamefont {J.~J.}\ \bibnamefont
  {Baumberg}}, \bibinfo {author} {\bibfnamefont {A.~V.}\ \bibnamefont
  {Kavokin}}, \bibinfo {author} {\bibfnamefont {S.}~\bibnamefont
  {Christopoulos}}, \bibinfo {author} {\bibfnamefont {A.~J.~D.}\ \bibnamefont
  {Grundy}}, \bibinfo {author} {\bibfnamefont {R.}~\bibnamefont {Butt\'e}},
  \bibinfo {author} {\bibfnamefont {G.}~\bibnamefont {Christmann}}, \bibinfo
  {author} {\bibfnamefont {D.~D.}\ \bibnamefont {Solnyshkov}}, \bibinfo
  {author} {\bibfnamefont {G.}~\bibnamefont {Malpuech}}, \bibinfo {author}
  {\bibfnamefont {G.}~\bibnamefont {Baldassarri H\"oger~von H\"ogersthal}},
  \bibinfo {author} {\bibfnamefont {E.}~\bibnamefont {Feltin}}, \bibinfo
  {author} {\bibfnamefont {J.-F.}\ \bibnamefont {Carlin}}, \ and\ \bibinfo
  {author} {\bibfnamefont {N.}~\bibnamefont {Grandjean}},\ }\href {\doibase
  10.1103/PhysRevLett.101.136409} {\bibfield  {journal} {\bibinfo  {journal}
  {Phys. Rev. Lett.}\ }\textbf {\bibinfo {volume} {101}},\ \bibinfo {pages}
  {136409} (\bibinfo {year} {2008})}\BibitemShut {NoStop}%
\bibitem [{\citenamefont {Wertz}\ \emph {et~al.}(2010)\citenamefont {Wertz},
  \citenamefont {Ferrier}, \citenamefont {Solnyshkov}, \citenamefont {Johne},
  \citenamefont {Sanvitto}, \citenamefont {Lemaitre}, \citenamefont {Sagnes},
  \citenamefont {Grousson}, \citenamefont {Kavokin}, \citenamefont {Senellart},
  \citenamefont {Malpuech},\ and\ \citenamefont {Bloch}}]{Wertz2010}%
  \BibitemOpen
  \bibfield  {author} {\bibinfo {author} {\bibfnamefont {E.}~\bibnamefont
  {Wertz}}, \bibinfo {author} {\bibfnamefont {L.}~\bibnamefont {Ferrier}},
  \bibinfo {author} {\bibfnamefont {D.~D.}\ \bibnamefont {Solnyshkov}},
  \bibinfo {author} {\bibfnamefont {R.}~\bibnamefont {Johne}}, \bibinfo
  {author} {\bibfnamefont {D.}~\bibnamefont {Sanvitto}}, \bibinfo {author}
  {\bibfnamefont {A.}~\bibnamefont {Lemaitre}}, \bibinfo {author}
  {\bibfnamefont {I.}~\bibnamefont {Sagnes}}, \bibinfo {author} {\bibfnamefont
  {R.}~\bibnamefont {Grousson}}, \bibinfo {author} {\bibfnamefont {A.~V.}\
  \bibnamefont {Kavokin}}, \bibinfo {author} {\bibfnamefont {P.}~\bibnamefont
  {Senellart}}, \bibinfo {author} {\bibfnamefont {G.}~\bibnamefont {Malpuech}},
  \ and\ \bibinfo {author} {\bibfnamefont {J.}~\bibnamefont {Bloch}},\ }\href
  {http://dx.doi.org/10.1038/nphys1750} {\bibfield  {journal} {\bibinfo
  {journal} {Nat Phys}\ }\textbf {\bibinfo {volume} {6}},\ \bibinfo {pages}
  {860} (\bibinfo {year} {2010})}\BibitemShut {NoStop}%
\bibitem [{\citenamefont {Snoke}(2002)}]{Snoke2002}%
  \BibitemOpen
  \bibfield  {author} {\bibinfo {author} {\bibfnamefont {D.}~\bibnamefont
  {Snoke}},\ }\href {\doibase 10.1126/science.1078082} {\bibfield  {journal}
  {\bibinfo  {journal} {Science}\ }\textbf {\bibinfo {volume} {298}},\ \bibinfo
  {pages} {1368} (\bibinfo {year} {2002})}\BibitemShut {NoStop}%
\bibitem [{\citenamefont {Deng}\ \emph {et~al.}(2010)\citenamefont {Deng},
  \citenamefont {Haug},\ and\ \citenamefont {Yamamoto}}]{Deng2010}%
  \BibitemOpen
  \bibfield  {author} {\bibinfo {author} {\bibfnamefont {H.}~\bibnamefont
  {Deng}}, \bibinfo {author} {\bibfnamefont {H.}~\bibnamefont {Haug}}, \ and\
  \bibinfo {author} {\bibfnamefont {Y.}~\bibnamefont {Yamamoto}},\ }\href
  {\doibase 10.1103/RevModPhys.82.1489} {\bibfield  {journal} {\bibinfo
  {journal} {Rev. Mod. Phys.}\ }\textbf {\bibinfo {volume} {82}},\ \bibinfo
  {pages} {1489} (\bibinfo {year} {2010})}\BibitemShut {NoStop}%
\bibitem [{\citenamefont {Carusotto}\ and\ \citenamefont
  {Ciuti}(2013)}]{Carusotto2013}%
  \BibitemOpen
  \bibfield  {author} {\bibinfo {author} {\bibfnamefont {I.}~\bibnamefont
  {Carusotto}}\ and\ \bibinfo {author} {\bibfnamefont {C.}~\bibnamefont
  {Ciuti}},\ }\href {\doibase 10.1103/RevModPhys.85.299} {\bibfield  {journal}
  {\bibinfo  {journal} {Rev. Mod. Phys.}\ }\textbf {\bibinfo {volume} {85}},\
  \bibinfo {pages} {299} (\bibinfo {year} {2013})}\BibitemShut {NoStop}%
\bibitem [{\citenamefont {Byrnes}\ \emph {et~al.}(2014)\citenamefont {Byrnes},
  \citenamefont {Kim},\ and\ \citenamefont {Yamamoto}}]{Byrnes2014}%
  \BibitemOpen
  \bibfield  {author} {\bibinfo {author} {\bibfnamefont {T.}~\bibnamefont
  {Byrnes}}, \bibinfo {author} {\bibfnamefont {N.~Y.}\ \bibnamefont {Kim}}, \
  and\ \bibinfo {author} {\bibfnamefont {Y.}~\bibnamefont {Yamamoto}},\ }\href
  {http://dx.doi.org/10.1038/nphys3143} {\bibfield  {journal} {\bibinfo
  {journal} {Nat Phys}\ }\textbf {\bibinfo {volume} {10}},\ \bibinfo {pages}
  {803} (\bibinfo {year} {2014})}\BibitemShut {NoStop}%
\bibitem [{\citenamefont {Sanvitto}\ and\ \citenamefont
  {Kena-Cohen}(2016)}]{Sanvitto2016}%
  \BibitemOpen
  \bibfield  {author} {\bibinfo {author} {\bibfnamefont {D.}~\bibnamefont
  {Sanvitto}}\ and\ \bibinfo {author} {\bibfnamefont {S.}~\bibnamefont
  {Kena-Cohen}},\ }\href {http://dx.doi.org/10.1038/nmat4668} {\bibfield
  {journal} {\bibinfo  {journal} {Nat Mater}\ }\textbf {\bibinfo {volume}
  {15}},\ \bibinfo {pages} {1061} (\bibinfo {year} {2016})}\BibitemShut
  {NoStop}%
\bibitem [{\citenamefont {Sun}\ \emph {et~al.}(2017)\citenamefont {Sun},
  \citenamefont {Wen}, \citenamefont {Yoon}, \citenamefont {Liu}, \citenamefont
  {Steger}, \citenamefont {Pfeiffer}, \citenamefont {West}, \citenamefont
  {Snoke},\ and\ \citenamefont {Nelson}}]{Sun2017}%
  \BibitemOpen
  \bibfield  {author} {\bibinfo {author} {\bibfnamefont {Y.}~\bibnamefont
  {Sun}}, \bibinfo {author} {\bibfnamefont {P.}~\bibnamefont {Wen}}, \bibinfo
  {author} {\bibfnamefont {Y.}~\bibnamefont {Yoon}}, \bibinfo {author}
  {\bibfnamefont {G.}~\bibnamefont {Liu}}, \bibinfo {author} {\bibfnamefont
  {M.}~\bibnamefont {Steger}}, \bibinfo {author} {\bibfnamefont {L.~N.}\
  \bibnamefont {Pfeiffer}}, \bibinfo {author} {\bibfnamefont {K.}~\bibnamefont
  {West}}, \bibinfo {author} {\bibfnamefont {D.~W.}\ \bibnamefont {Snoke}}, \
  and\ \bibinfo {author} {\bibfnamefont {K.~A.}\ \bibnamefont {Nelson}},\
  }\href {\doibase 10.1103/PhysRevLett.118.016602} {\bibfield  {journal}
  {\bibinfo  {journal} {Phys. Rev. Lett.}\ }\textbf {\bibinfo {volume} {118}},\
  \bibinfo {pages} {016602} (\bibinfo {year} {2017})}\BibitemShut {NoStop}%
\bibitem [{\citenamefont {Schneider}\ \emph {et~al.}(2013)\citenamefont
  {Schneider}, \citenamefont {Rahimi-Iman}, \citenamefont {Kim}, \citenamefont
  {Fischer}, \citenamefont {Savenko}, \citenamefont {Amthor}, \citenamefont
  {Lermer}, \citenamefont {Wolf}, \citenamefont {Worschech}, \citenamefont
  {Kulakovskii}, \citenamefont {Shelykh}, \citenamefont {Kamp}, \citenamefont
  {Reitzenstein}, \citenamefont {Forchel}, \citenamefont {Yamamoto},\ and\
  \citenamefont {Hofling}}]{Schneider2013}%
  \BibitemOpen
  \bibfield  {author} {\bibinfo {author} {\bibfnamefont {C.}~\bibnamefont
  {Schneider}}, \bibinfo {author} {\bibfnamefont {A.}~\bibnamefont
  {Rahimi-Iman}}, \bibinfo {author} {\bibfnamefont {N.~Y.}\ \bibnamefont
  {Kim}}, \bibinfo {author} {\bibfnamefont {J.}~\bibnamefont {Fischer}},
  \bibinfo {author} {\bibfnamefont {I.~G.}\ \bibnamefont {Savenko}}, \bibinfo
  {author} {\bibfnamefont {M.}~\bibnamefont {Amthor}}, \bibinfo {author}
  {\bibfnamefont {M.}~\bibnamefont {Lermer}}, \bibinfo {author} {\bibfnamefont
  {A.}~\bibnamefont {Wolf}}, \bibinfo {author} {\bibfnamefont {L.}~\bibnamefont
  {Worschech}}, \bibinfo {author} {\bibfnamefont {V.~D.}\ \bibnamefont
  {Kulakovskii}}, \bibinfo {author} {\bibfnamefont {I.~A.}\ \bibnamefont
  {Shelykh}}, \bibinfo {author} {\bibfnamefont {M.}~\bibnamefont {Kamp}},
  \bibinfo {author} {\bibfnamefont {S.}~\bibnamefont {Reitzenstein}}, \bibinfo
  {author} {\bibfnamefont {A.}~\bibnamefont {Forchel}}, \bibinfo {author}
  {\bibfnamefont {Y.}~\bibnamefont {Yamamoto}}, \ and\ \bibinfo {author}
  {\bibfnamefont {S.}~\bibnamefont {Hofling}},\ }\href
  {http://dx.doi.org/10.1038/nature12036} {\bibfield  {journal} {\bibinfo
  {journal} {Nature}\ }\textbf {\bibinfo {volume} {497}},\ \bibinfo {pages}
  {348} (\bibinfo {year} {2013})}\BibitemShut {NoStop}%
\bibitem [{\citenamefont {Bhattacharya}\ \emph {et~al.}(2013)\citenamefont
  {Bhattacharya}, \citenamefont {Xiao}, \citenamefont {Das}, \citenamefont
  {Bhowmick},\ and\ \citenamefont {Heo}}]{Bhattacharya2013}%
  \BibitemOpen
  \bibfield  {author} {\bibinfo {author} {\bibfnamefont {P.}~\bibnamefont
  {Bhattacharya}}, \bibinfo {author} {\bibfnamefont {B.}~\bibnamefont {Xiao}},
  \bibinfo {author} {\bibfnamefont {A.}~\bibnamefont {Das}}, \bibinfo {author}
  {\bibfnamefont {S.}~\bibnamefont {Bhowmick}}, \ and\ \bibinfo {author}
  {\bibfnamefont {J.}~\bibnamefont {Heo}},\ }\href {\doibase
  10.1103/PhysRevLett.110.206403} {\bibfield  {journal} {\bibinfo  {journal}
  {Phys. Rev. Lett.}\ }\textbf {\bibinfo {volume} {110}},\ \bibinfo {pages}
  {206403} (\bibinfo {year} {2013})}\BibitemShut {NoStop}%
\bibitem [{\citenamefont {Bhattacharya}\ \emph {et~al.}(2014)\citenamefont
  {Bhattacharya}, \citenamefont {Frost}, \citenamefont {Deshpande},
  \citenamefont {Baten}, \citenamefont {Hazari},\ and\ \citenamefont
  {Das}}]{Bhattacharya2014}%
  \BibitemOpen
  \bibfield  {author} {\bibinfo {author} {\bibfnamefont {P.}~\bibnamefont
  {Bhattacharya}}, \bibinfo {author} {\bibfnamefont {T.}~\bibnamefont {Frost}},
  \bibinfo {author} {\bibfnamefont {S.}~\bibnamefont {Deshpande}}, \bibinfo
  {author} {\bibfnamefont {M.~Z.}\ \bibnamefont {Baten}}, \bibinfo {author}
  {\bibfnamefont {A.}~\bibnamefont {Hazari}}, \ and\ \bibinfo {author}
  {\bibfnamefont {A.}~\bibnamefont {Das}},\ }\href {\doibase
  10.1103/PhysRevLett.112.236802} {\bibfield  {journal} {\bibinfo  {journal}
  {Phys. Rev. Lett.}\ }\textbf {\bibinfo {volume} {112}},\ \bibinfo {pages}
  {236802} (\bibinfo {year} {2014})}\BibitemShut {NoStop}%
\bibitem [{\citenamefont {Yao}\ \emph {et~al.}(2012)\citenamefont {Yao},
  \citenamefont {Hoffman},\ and\ \citenamefont {Gmachl}}]{Yao2012}%
  \BibitemOpen
  \bibfield  {author} {\bibinfo {author} {\bibfnamefont {Y.}~\bibnamefont
  {Yao}}, \bibinfo {author} {\bibfnamefont {A.~J.}\ \bibnamefont {Hoffman}}, \
  and\ \bibinfo {author} {\bibfnamefont {C.~F.}\ \bibnamefont {Gmachl}},\
  }\href {http://dx.doi.org/10.1038/nphoton.2012.143} {\bibfield  {journal}
  {\bibinfo  {journal} {Nat. Photonics}\ }\textbf {\bibinfo {volume} {6}},\
  \bibinfo {pages} {432} (\bibinfo {year} {2012})}\BibitemShut {NoStop}%
\bibitem [{Note1()}]{Note1}%
  \BibitemOpen
  \bibinfo {note} {For systems in which dark excitons (excitons with momentum
  or spin quantum numbers that don't match those of the photons) are favored
  however, direct optical access is blocked.}\BibitemShut {Stop}%
\bibitem [{\citenamefont {Cohen}\ \emph {et~al.}(2016)\citenamefont {Cohen},
  \citenamefont {Shilo}, \citenamefont {West}, \citenamefont {Pfeiffer},\ and\
  \citenamefont {Rapaport}}]{Cohen2016}%
  \BibitemOpen
  \bibfield  {author} {\bibinfo {author} {\bibfnamefont {K.}~\bibnamefont
  {Cohen}}, \bibinfo {author} {\bibfnamefont {Y.}~\bibnamefont {Shilo}},
  \bibinfo {author} {\bibfnamefont {K.}~\bibnamefont {West}}, \bibinfo {author}
  {\bibfnamefont {L.}~\bibnamefont {Pfeiffer}}, \ and\ \bibinfo {author}
  {\bibfnamefont {R.}~\bibnamefont {Rapaport}},\ }\href {\doibase
  10.1021/acs.nanolett.6b01061} {\bibfield  {journal} {\bibinfo  {journal}
  {Nano Lett.}\ }\textbf {\bibinfo {volume} {16}},\ \bibinfo {pages} {3726}
  (\bibinfo {year} {2016})}\BibitemShut {NoStop}%
\bibitem [{\citenamefont {Shilo}\ \emph {et~al.}(2013)\citenamefont {Shilo},
  \citenamefont {Cohen}, \citenamefont {Laikhtman}, \citenamefont {West},
  \citenamefont {Pfeiffer},\ and\ \citenamefont {Rapaport}}]{Shilo2013}%
  \BibitemOpen
  \bibfield  {author} {\bibinfo {author} {\bibfnamefont {Y.}~\bibnamefont
  {Shilo}}, \bibinfo {author} {\bibfnamefont {K.}~\bibnamefont {Cohen}},
  \bibinfo {author} {\bibfnamefont {B.}~\bibnamefont {Laikhtman}}, \bibinfo
  {author} {\bibfnamefont {K.}~\bibnamefont {West}}, \bibinfo {author}
  {\bibfnamefont {L.}~\bibnamefont {Pfeiffer}}, \ and\ \bibinfo {author}
  {\bibfnamefont {R.}~\bibnamefont {Rapaport}},\ }\href
  {http://dx.doi.org/10.1038/ncomms3335} {\bibfield  {journal} {\bibinfo
  {journal} {Nat. Comm.}\ }\textbf {\bibinfo {volume} {4}},\ \bibinfo {pages}
  {2335 EP} (\bibinfo {year} {2013})}\BibitemShut {NoStop}%
\bibitem [{\citenamefont {Qiu}\ \emph {et~al.}(2013)\citenamefont {Qiu},
  \citenamefont {da~Jornada},\ and\ \citenamefont {Louie}}]{Qiu2013}%
  \BibitemOpen
  \bibfield  {author} {\bibinfo {author} {\bibfnamefont {D.~Y.}\ \bibnamefont
  {Qiu}}, \bibinfo {author} {\bibfnamefont {F.~H.}\ \bibnamefont {da~Jornada}},
  \ and\ \bibinfo {author} {\bibfnamefont {S.~G.}\ \bibnamefont {Louie}},\
  }\href {\doibase 10.1103/PhysRevLett.111.216805} {\bibfield  {journal}
  {\bibinfo  {journal} {Phys. Rev. Lett.}\ }\textbf {\bibinfo {volume} {111}},\
  \bibinfo {pages} {216805} (\bibinfo {year} {2013})}\BibitemShut {NoStop}%
\bibitem [{\citenamefont {Mak}\ \emph {et~al.}(2013)\citenamefont {Mak},
  \citenamefont {He}, \citenamefont {Lee}, \citenamefont {Lee}, \citenamefont
  {Hone}, \citenamefont {Heinz},\ and\ \citenamefont {Shan}}]{Mak2013}%
  \BibitemOpen
  \bibfield  {author} {\bibinfo {author} {\bibfnamefont {K.~F.}\ \bibnamefont
  {Mak}}, \bibinfo {author} {\bibfnamefont {K.}~\bibnamefont {He}}, \bibinfo
  {author} {\bibfnamefont {C.}~\bibnamefont {Lee}}, \bibinfo {author}
  {\bibfnamefont {G.~H.}\ \bibnamefont {Lee}}, \bibinfo {author} {\bibfnamefont
  {J.}~\bibnamefont {Hone}}, \bibinfo {author} {\bibfnamefont {T.~F.}\
  \bibnamefont {Heinz}}, \ and\ \bibinfo {author} {\bibfnamefont
  {J.}~\bibnamefont {Shan}},\ }\href {http://dx.doi.org/10.1038/nmat3505}
  {\bibfield  {journal} {\bibinfo  {journal} {Nat. Mater.}\ }\textbf {\bibinfo
  {volume} {12}},\ \bibinfo {pages} {207} (\bibinfo {year} {2013})}\BibitemShut
  {NoStop}%
\bibitem [{\citenamefont {Fogler}\ \emph {et~al.}(2014)\citenamefont {Fogler},
  \citenamefont {Butov},\ and\ \citenamefont {Novoselov}}]{Fogler2014}%
  \BibitemOpen
  \bibfield  {author} {\bibinfo {author} {\bibfnamefont {M.~M.}\ \bibnamefont
  {Fogler}}, \bibinfo {author} {\bibfnamefont {L.~V.}\ \bibnamefont {Butov}}, \
  and\ \bibinfo {author} {\bibfnamefont {K.~S.}\ \bibnamefont {Novoselov}},\
  }\href {http://dx.doi.org/10.1038/ncomms5555} {\bibfield  {journal} {\bibinfo
   {journal} {Nat. Comm.}\ }\textbf {\bibinfo {volume} {5}},\ \bibinfo {pages}
  {4555 EP} (\bibinfo {year} {2014})}\BibitemShut {NoStop}%
\bibitem [{\citenamefont {Geim}\ and\ \citenamefont
  {Grigorieva}(2013)}]{Geim2013}%
  \BibitemOpen
  \bibfield  {author} {\bibinfo {author} {\bibfnamefont {A.~K.}\ \bibnamefont
  {Geim}}\ and\ \bibinfo {author} {\bibfnamefont {I.~V.}\ \bibnamefont
  {Grigorieva}},\ }\href {http://dx.doi.org/10.1038/nature12385} {\bibfield
  {journal} {\bibinfo  {journal} {Nature}\ }\textbf {\bibinfo {volume} {499}},\
  \bibinfo {pages} {419} (\bibinfo {year} {2013})}\BibitemShut {NoStop}%
\bibitem [{\citenamefont {Novoselov}\ \emph {et~al.}(2016)\citenamefont
  {Novoselov}, \citenamefont {Mishchenko}, \citenamefont {Carvalho},\ and\
  \citenamefont {Castro~Neto}}]{Novoselov2016}%
  \BibitemOpen
  \bibfield  {author} {\bibinfo {author} {\bibfnamefont {K.~S.}\ \bibnamefont
  {Novoselov}}, \bibinfo {author} {\bibfnamefont {A.}~\bibnamefont
  {Mishchenko}}, \bibinfo {author} {\bibfnamefont {A.}~\bibnamefont
  {Carvalho}}, \ and\ \bibinfo {author} {\bibfnamefont {A.~H.}\ \bibnamefont
  {Castro~Neto}},\ }\href
  {http://science.sciencemag.org/content/353/6298/aac9439} {\bibfield
  {journal} {\bibinfo  {journal} {Science}\ }\textbf {\bibinfo {volume}
  {353}},\ \bibinfo {pages} {aac9439} (\bibinfo {year} {2016})}\BibitemShut
  {NoStop}%
\bibitem [{\citenamefont {Mak}\ and\ \citenamefont {Shan}(2016)}]{Mak2016}%
  \BibitemOpen
  \bibfield  {author} {\bibinfo {author} {\bibfnamefont {K.~F.}\ \bibnamefont
  {Mak}}\ and\ \bibinfo {author} {\bibfnamefont {J.}~\bibnamefont {Shan}},\
  }\href {http://dx.doi.org/10.1038/nphoton.2015.282} {\bibfield  {journal}
  {\bibinfo  {journal} {Nat. Photonics}\ }\textbf {\bibinfo {volume} {10}},\
  \bibinfo {pages} {216} (\bibinfo {year} {2016})}\BibitemShut {NoStop}%
\bibitem [{\citenamefont {Rivera}\ \emph {et~al.}(2015)\citenamefont {Rivera},
  \citenamefont {Schaibley}, \citenamefont {Jones}, \citenamefont {Ross},
  \citenamefont {Wu}, \citenamefont {Aivazian}, \citenamefont {Klement},
  \citenamefont {Seyler}, \citenamefont {Clark}, \citenamefont {Ghimire},
  \citenamefont {Yan}, \citenamefont {Mandrus}, \citenamefont {Yao},\ and\
  \citenamefont {Xu}}]{Rivera2015}%
  \BibitemOpen
  \bibfield  {author} {\bibinfo {author} {\bibfnamefont {P.}~\bibnamefont
  {Rivera}}, \bibinfo {author} {\bibfnamefont {J.~R.}\ \bibnamefont
  {Schaibley}}, \bibinfo {author} {\bibfnamefont {A.~M.}\ \bibnamefont
  {Jones}}, \bibinfo {author} {\bibfnamefont {J.~S.}\ \bibnamefont {Ross}},
  \bibinfo {author} {\bibfnamefont {S.}~\bibnamefont {Wu}}, \bibinfo {author}
  {\bibfnamefont {G.}~\bibnamefont {Aivazian}}, \bibinfo {author}
  {\bibfnamefont {P.}~\bibnamefont {Klement}}, \bibinfo {author} {\bibfnamefont
  {K.}~\bibnamefont {Seyler}}, \bibinfo {author} {\bibfnamefont
  {G.}~\bibnamefont {Clark}}, \bibinfo {author} {\bibfnamefont {N.~J.}\
  \bibnamefont {Ghimire}}, \bibinfo {author} {\bibfnamefont {J.}~\bibnamefont
  {Yan}}, \bibinfo {author} {\bibfnamefont {D.~G.}\ \bibnamefont {Mandrus}},
  \bibinfo {author} {\bibfnamefont {W.}~\bibnamefont {Yao}}, \ and\ \bibinfo
  {author} {\bibfnamefont {X.}~\bibnamefont {Xu}},\ }\href
  {http://dx.doi.org/10.1038/ncomms7242} {\bibfield  {journal} {\bibinfo
  {journal} {Nat. Commun.}\ }\textbf {\bibinfo {volume} {6}},\ \bibinfo {pages}
  {6242} (\bibinfo {year} {2015})}\BibitemShut {NoStop}%
\bibitem [{\citenamefont {Palummo}\ \emph {et~al.}(2015)\citenamefont
  {Palummo}, \citenamefont {Bernardi},\ and\ \citenamefont
  {Grossman}}]{Palummo2015}%
  \BibitemOpen
  \bibfield  {author} {\bibinfo {author} {\bibfnamefont {M.}~\bibnamefont
  {Palummo}}, \bibinfo {author} {\bibfnamefont {M.}~\bibnamefont {Bernardi}}, \
  and\ \bibinfo {author} {\bibfnamefont {J.~C.}\ \bibnamefont {Grossman}},\
  }\href {\doibase 10.1021/nl503799t} {\bibfield  {journal} {\bibinfo
  {journal} {Nano Lett.}\ }\textbf {\bibinfo {volume} {15}},\ \bibinfo {pages}
  {2794} (\bibinfo {year} {2015})}\BibitemShut {NoStop}%
\bibitem [{\citenamefont {{Calman}}\ \emph {et~al.}()\citenamefont {{Calman}},
  \citenamefont {{Fogler}}, \citenamefont {{Butov}}, \citenamefont {{Hu}},
  \citenamefont {{Mishchenko}},\ and\ \citenamefont {{Geim}}}]{Calman2017}%
  \BibitemOpen
  \bibfield  {author} {\bibinfo {author} {\bibfnamefont {E.~V.}\ \bibnamefont
  {{Calman}}}, \bibinfo {author} {\bibfnamefont {M.~M.}\ \bibnamefont
  {{Fogler}}}, \bibinfo {author} {\bibfnamefont {L.~V.}\ \bibnamefont
  {{Butov}}}, \bibinfo {author} {\bibfnamefont {S.}~\bibnamefont {{Hu}}},
  \bibinfo {author} {\bibfnamefont {A.}~\bibnamefont {{Mishchenko}}}, \ and\
  \bibinfo {author} {\bibfnamefont {A.~K.}\ \bibnamefont {{Geim}}},\
  }\href@noop {} {\ }\Eprint {http://arxiv.org/abs/1709.07043}
  {arXiv:1709.07043} \BibitemShut {NoStop}%
\bibitem [{\citenamefont {Raja}\ \emph {et~al.}(2017)\citenamefont {Raja},
  \citenamefont {Chaves}, \citenamefont {Yu}, \citenamefont {Arefe},
  \citenamefont {Hill}, \citenamefont {Rigosi}, \citenamefont {Berkelbach},
  \citenamefont {Nagler}, \citenamefont {Sch{\"u}ller}, \citenamefont {Korn},
  \citenamefont {Nuckolls}, \citenamefont {Hone}, \citenamefont {Brus},
  \citenamefont {Heinz}, \citenamefont {Reichman},\ and\ \citenamefont
  {Chernikov}}]{raja2017}%
  \BibitemOpen
  \bibfield  {author} {\bibinfo {author} {\bibfnamefont {A.}~\bibnamefont
  {Raja}}, \bibinfo {author} {\bibfnamefont {A.}~\bibnamefont {Chaves}},
  \bibinfo {author} {\bibfnamefont {J.}~\bibnamefont {Yu}}, \bibinfo {author}
  {\bibfnamefont {G.}~\bibnamefont {Arefe}}, \bibinfo {author} {\bibfnamefont
  {H.~M.}\ \bibnamefont {Hill}}, \bibinfo {author} {\bibfnamefont {A.~F.}\
  \bibnamefont {Rigosi}}, \bibinfo {author} {\bibfnamefont {T.~C.}\
  \bibnamefont {Berkelbach}}, \bibinfo {author} {\bibfnamefont
  {P.}~\bibnamefont {Nagler}}, \bibinfo {author} {\bibfnamefont
  {C.}~\bibnamefont {Sch{\"u}ller}}, \bibinfo {author} {\bibfnamefont
  {T.}~\bibnamefont {Korn}}, \bibinfo {author} {\bibfnamefont {C.}~\bibnamefont
  {Nuckolls}}, \bibinfo {author} {\bibfnamefont {J.}~\bibnamefont {Hone}},
  \bibinfo {author} {\bibfnamefont {L.~E.}\ \bibnamefont {Brus}}, \bibinfo
  {author} {\bibfnamefont {T.~F.}\ \bibnamefont {Heinz}}, \bibinfo {author}
  {\bibfnamefont {D.~R.}\ \bibnamefont {Reichman}}, \ and\ \bibinfo {author}
  {\bibfnamefont {A.}~\bibnamefont {Chernikov}},\ }\href
  {http://dx.doi.org/10.1038/ncomms15251} {\bibfield  {journal} {\bibinfo
  {journal} {Nat. Comm.}\ }\textbf {\bibinfo {volume} {8}},\ \bibinfo {pages}
  {15251 EP } (\bibinfo {year} {2017})}\BibitemShut {NoStop}%
\bibitem [{\citenamefont {Zhang}\ \emph {et~al.}(2017)\citenamefont {Zhang},
  \citenamefont {Gong}, \citenamefont {Nie}, \citenamefont {Min}, \citenamefont
  {Liang}, \citenamefont {Oh}, \citenamefont {Zhang}, \citenamefont {Wang},
  \citenamefont {Hong}, \citenamefont {Colombo}, \citenamefont {Wallace},\ and\
  \citenamefont {Cho}}]{Zhang2017}%
  \BibitemOpen
  \bibfield  {author} {\bibinfo {author} {\bibfnamefont {C.}~\bibnamefont
  {Zhang}}, \bibinfo {author} {\bibfnamefont {C.}~\bibnamefont {Gong}},
  \bibinfo {author} {\bibfnamefont {Y.}~\bibnamefont {Nie}}, \bibinfo {author}
  {\bibfnamefont {K.-A.}\ \bibnamefont {Min}}, \bibinfo {author} {\bibfnamefont
  {C.}~\bibnamefont {Liang}}, \bibinfo {author} {\bibfnamefont {Y.~J.}\
  \bibnamefont {Oh}}, \bibinfo {author} {\bibfnamefont {H.}~\bibnamefont
  {Zhang}}, \bibinfo {author} {\bibfnamefont {W.}~\bibnamefont {Wang}},
  \bibinfo {author} {\bibfnamefont {S.}~\bibnamefont {Hong}}, \bibinfo {author}
  {\bibfnamefont {L.}~\bibnamefont {Colombo}}, \bibinfo {author} {\bibfnamefont
  {R.~M.}\ \bibnamefont {Wallace}}, \ and\ \bibinfo {author} {\bibfnamefont
  {K.}~\bibnamefont {Cho}},\ }\href
  {http://stacks.iop.org/2053-1583/4/i=1/a=015026} {\bibfield  {journal}
  {\bibinfo  {journal} {2D Materials}\ }\textbf {\bibinfo {volume} {4}},\
  \bibinfo {pages} {015026} (\bibinfo {year} {2017})}\BibitemShut {NoStop}%
\bibitem [{\citenamefont {Bistritzer}\ and\ \citenamefont
  {MacDonald}(2010)}]{Bistritzer2010}%
  \BibitemOpen
  \bibfield  {author} {\bibinfo {author} {\bibfnamefont {R.}~\bibnamefont
  {Bistritzer}}\ and\ \bibinfo {author} {\bibfnamefont {A.~H.}\ \bibnamefont
  {MacDonald}},\ }\href {\doibase 10.1103/PhysRevB.81.245412} {\bibfield
  {journal} {\bibinfo  {journal} {Phys. Rev. B}\ }\textbf {\bibinfo {volume}
  {81}},\ \bibinfo {pages} {245412} (\bibinfo {year} {2010})}\BibitemShut
  {NoStop}%
\bibitem [{\citenamefont {Zhou}\ \emph {et~al.}(2017)\citenamefont {Zhou},
  \citenamefont {Wickramaratne}, \citenamefont {Ge}, \citenamefont {Su},
  \citenamefont {De},\ and\ \citenamefont {Lake}}]{Zhou2017}%
  \BibitemOpen
  \bibfield  {author} {\bibinfo {author} {\bibfnamefont {K.}~\bibnamefont
  {Zhou}}, \bibinfo {author} {\bibfnamefont {D.}~\bibnamefont {Wickramaratne}},
  \bibinfo {author} {\bibfnamefont {S.}~\bibnamefont {Ge}}, \bibinfo {author}
  {\bibfnamefont {S.}~\bibnamefont {Su}}, \bibinfo {author} {\bibfnamefont
  {A.}~\bibnamefont {De}}, \ and\ \bibinfo {author} {\bibfnamefont {R.~K.}\
  \bibnamefont {Lake}},\ }\href {\doibase 10.1039/C6CP08927E} {\bibfield
  {journal} {\bibinfo  {journal} {Phys. Chem. Chem. Phys.}\ }\textbf {\bibinfo
  {volume} {19}},\ \bibinfo {pages} {10406} (\bibinfo {year}
  {2017})}\BibitemShut {NoStop}%
\bibitem [{\citenamefont {Comte}\ and\ \citenamefont
  {Nozières}(1982)}]{Comte1982}%
  \BibitemOpen
  \bibfield  {author} {\bibinfo {author} {\bibfnamefont {C.}~\bibnamefont
  {Comte}}\ and\ \bibinfo {author} {\bibfnamefont {P.}~\bibnamefont
  {Nozières}},\ }\href {\doibase 10.1051/jphys:019820043070106900} {\bibfield
  {journal} {\bibinfo  {journal} {J. Phys. France}\ }\textbf {\bibinfo {volume}
  {43}},\ \bibinfo {pages} {1069} (\bibinfo {year} {1982})}\BibitemShut
  {NoStop}%
\bibitem [{\citenamefont {Zhu}\ \emph {et~al.}(1995)\citenamefont {Zhu},
  \citenamefont {Littlewood}, \citenamefont {Hybertsen},\ and\ \citenamefont
  {Rice}}]{Zhu1995}%
  \BibitemOpen
  \bibfield  {author} {\bibinfo {author} {\bibfnamefont {X.}~\bibnamefont
  {Zhu}}, \bibinfo {author} {\bibfnamefont {P.~B.}\ \bibnamefont {Littlewood}},
  \bibinfo {author} {\bibfnamefont {M.~S.}\ \bibnamefont {Hybertsen}}, \ and\
  \bibinfo {author} {\bibfnamefont {T.~M.}\ \bibnamefont {Rice}},\ }\href
  {\doibase 10.1103/PhysRevLett.74.1633} {\bibfield  {journal} {\bibinfo
  {journal} {Phys. Rev. Lett.}\ }\textbf {\bibinfo {volume} {74}},\ \bibinfo
  {pages} {1633} (\bibinfo {year} {1995})}\BibitemShut {NoStop}%
\bibitem [{\citenamefont {Wu}\ \emph {et~al.}(2015)\citenamefont {Wu},
  \citenamefont {Xue},\ and\ \citenamefont {MacDonald}}]{Wu2015}%
  \BibitemOpen
  \bibfield  {author} {\bibinfo {author} {\bibfnamefont {F.-C.}\ \bibnamefont
  {Wu}}, \bibinfo {author} {\bibfnamefont {F.}~\bibnamefont {Xue}}, \ and\
  \bibinfo {author} {\bibfnamefont {A.~H.}\ \bibnamefont {MacDonald}},\ }\href
  {\doibase 10.1103/PhysRevB.92.165121} {\bibfield  {journal} {\bibinfo
  {journal} {Phys. Rev. B}\ }\textbf {\bibinfo {volume} {92}},\ \bibinfo
  {pages} {165121} (\bibinfo {year} {2015})}\BibitemShut {NoStop}%
\bibitem [{Note17()}]{Note17}%
  \BibitemOpen
  \bibinfo {note} {See Supplemental Material for the discussion of the total
  energy functional $\protect \mathcal {E}[n_S,n_{ex}]$, the existence of a
  global minimum and a detailed derivation of the {\protect \it ac} response
  equations.}\BibitemShut {Stop}%
\bibitem [{\citenamefont {Xia}\ \emph {et~al.}(2009)\citenamefont {Xia},
  \citenamefont {Chen}, \citenamefont {Li},\ and\ \citenamefont
  {Tao}}]{Xia2009}%
  \BibitemOpen
  \bibfield  {author} {\bibinfo {author} {\bibfnamefont {J.}~\bibnamefont
  {Xia}}, \bibinfo {author} {\bibfnamefont {F.}~\bibnamefont {Chen}}, \bibinfo
  {author} {\bibfnamefont {J.}~\bibnamefont {Li}}, \ and\ \bibinfo {author}
  {\bibfnamefont {N.}~\bibnamefont {Tao}},\ }\href
  {http://dx.doi.org/10.1038/nnano.2009.177} {\bibfield  {journal} {\bibinfo
  {journal} {Nat. Nanotechnol.}\ }\textbf {\bibinfo {volume} {4}},\ \bibinfo
  {pages} {505} (\bibinfo {year} {2009})}\BibitemShut {NoStop}%
\end{thebibliography}%
\bibliographystyle{apsrev4-1}

\end{document}